# Auxetic tetrahex-carbon with ultrahigh strength and direct band gap

Qun Wei,[1] Guang Yang,[2,3] Xihong Peng[2]

[1] School of Physics and Optoelectronic Engineering, Xidian University, Xi'an, Shaanxi 710071, P. R. China

[2] College of Integrative Sciences and Arts, Arizona State University, Mesa, Arizona 85212, USA

[3] School of Science, Hebei University of Science and Technology, Shijiazhuang, Hebei 050018, P. R. China

**ABSTRACT**

Tetrahex-carbon is a recently predicted two dimensional (2D) carbon allotrope which is composed of tetragonal and hexagonal rings. Unlike flat graphene, this new 2D carbon structure is buckled, possesses a direct band gap ~ 2.6 eV and high carrier mobility ~ $10^4$ $cm^2/$ (V·s) with anisotropic feature. In this work, we employ first-principles density-functional theory calculations to explore mechanical properties of tetrahex-C under uniaxial tensile strain. We find that tetrahex-C demonstrates ultrahigh ideal strength, outperforming both graphene and penta-graphene. It shows superior ductility and sustains uniaxial tensile strain up to 20% (16%) till phonon instability occurs, and the corresponding maximal strength is 38.3 N/m (37.8 N/m) in the zigzag (armchair) direction. It shows intrinsic negative Poisson's ratio. This exotic in-plane Poisson's ratio takes place when axial strain reaches a threshold value of 7% (5%) in the zigzag (armchair) direction. We also find that Tetrahex-C holds a direct band gap of 2.64 eV at the center of Brillouin zone. This direct-gap feature maintains intact upon strain application with no direct-indirect gap transition. The ultrahigh ideal strength, negative Poisson's ratio and integrity of direct-gap under strain in tetrahex-C suggest it may have potential applications in nanomechanics and nanoelectronics.

## 1. Introduction

Successful fabrication of two dimensional (2D) materials such as graphene [1–3], transition metal dichalcogenides (TMDs) [4–7], and phosphorene [8,9] triggers tremendous interests of materials research in 2D structures. Graphene has been considered as a promising material for many advanced applications in future electronics [1–3]. However the intrinsic zero band gap in graphene limits its applications in electronic devices. Many other 2D allotropes of carbon were proposed, for example, S-graphene [10], T-graphene [11], various phases of carbon sheets [12,13], and *α*-, *β*-, *δ*-graphynes [14,15]. Penta-graphene [16] composed of only carbon pentagons was theoretically predicted a few years ago and have drawn plenty of research attention due to its unique mechanical and electronic properties, such as high strength and quasi-direct band gap. Based on penta-graphene, Ram and Mizuseki [17] applied the Stone-Wales transformation [18] to generate a new 2D carbon allotrope which consists of tetragonal and hexagonal rings. This new structure tetrahex-C shows slightly lower energy than penta-graphene, implying a larger chance to be fabricated in reality. Fantastically, it shows a direct band gap of 2.6 eV with high electron mobility $\sim 10^4$ cm$^2$/(V·s) [17]. Carbon family such as diamond, graphene, carbon nanotubes has a reputation of excellent mechanical properties in terms of extraordinary strength and enormous moduli. It is of interest to evaluate the mechanical properties in this new tetrahex-C structure. We find that this material exhibits ultrahigh ideal strength outperforming both graphene and penta-graphene. Remarkably, it also demonstrates intrinsic negative Poisson's ratio.

Monolayer semiconducting TMDs (MX$_2$; M = Mo, W and X = S, Se, Te) and phosphorene possess a significant advantage over graphene in that they exhibit a direct band gap which is appropriate for applications in optoelectronic devices. However, these materials experience direct-to-indirect band gap transition upon strain application. For example, 2D MoS$_2$ and WS$_2$ undergoes direct-to-indirect band gap transition with a moderate uniaxial strain of ~ 2% [19–21]. MoSe$_2$ and WSe$_2$ sustain slightly higher strain to ~ 6% till the indirect-gap transition [20]. Phosphorene experiences direct-to-indirect band gap transition at 8% axial strain [22]. We find that tetrahex-C remains integrity of direct-gap within the entire range of strain application up 20% (16%) in the zigzag (armchair) direction.

## 2. Computational methods

The first-principles density-functional theory (DFT) [23] calculations are carried out using the Vienna *ab initio* simulation package (VASP) [24,25] with the projector-augmented wave (PAW) [26,27] potentials. The Perdew-Burke-Ernzerhof (PBE) exchange-correlation functional [28] is chosen for general electronic structure calculations and geometry relaxation. The hybrid Heyd-Scuseria-Ernzerhof (HSE)06 method [29,30] is used to calculate electronic band structures. Since the exchange-correlation functional in HSE method uses a mixing parameter to incorporate Hartree-Fock (HF) exact exchange and PBE functional, it has a better prediction on the band gap of semiconductors. In this study, the fraction of the HF exchange is set to be the default value of 0.25.

The wave functions of valence electrons are described using the plane wave basis set. The reciprocal space is meshed using Monkhorst-Pack method. The kinetic energy cutoff 900 eV for the basis set and $15 \times 13 \times 1$ mesh for reciprocal space are chosen in geometry relaxation and force fields calculations. The energy convergence criterion for electronic iterations is set to be $10^{-6}$ eV and the force is converged to $10^{-5}$ eV for geometry optimization of the unit cell. The kinetic energy cutoff 500 eV for plane wave basis set is used for HSE band structure calculations. In the band structure, 11 $k$-points are collected along each high symmetry line in the reciprocal space. The $z$-vector of the unit cell is set to be 20 Å to ensure sufficient vacuum space included in the calculations to minimize the interaction between the system and its replicas from the periodic boundary condition. The phonon frequencies are calculated using a supercell approach in the PHONOPY code [31] with the forces computed from VASP.

The initial structure of tetrahex-C is constructed according to the reference [17]. Unlike flat graphene, this 2D carbon network is buckled and composed with tetragonal and hexagonal rings as shown in Fig. 1. The unit cell contains 12 carbon atoms, which are either $sp^2$ or $sp^3$ hybridized at a ratio of 2:1. The $sp^3$ hybridized carbon is sandwiched between two layers of $sp^2$ bonded atoms.

Starting with the fully relaxed tetrahex-C structure, uniaxial tensile strain up to 40% at an increment of 1% is applied in either the $x$ (zigzag) or $y$ (armchair) direction to explore its strain-stress relation and determine ideal strength (the highest strength of a crystal at 0 K) [32,33] and critical strain (at which ideal strength reaches) [22]. The tensile strain is defined as,

$$\varepsilon = \frac{a-a_0}{a_0} \qquad (1)$$

where $a$ and $a_0$ are the lattice constants of the strained and relaxed structure, respectively. With strain applied in one direction, the lattice constant in the transvers direction is fully relaxed through minimization of the total energy to ensure no stress in the transverse direction. According to Equation (1), the response strain in the transverse direction can be also calculated.

Poisson's ratio is defined as,

$$v = -\frac{\varepsilon_{transverse}}{\varepsilon_{axial}}, v_{xy} = -\frac{\varepsilon_y}{\varepsilon_x}, v_{yx} = -\frac{\varepsilon_x}{\varepsilon_y} \quad (2)$$

where $\varepsilon_{axial}$ and $\varepsilon_{transverse}$ are the applied axial strain and its response strain in the transverse direction, respectively. In order to depict the nonlinear lattice response for finite strain, Poisson's ratio is usually calculated using finite difference method as [34–36],

$$v = -\frac{d\,\varepsilon_{transverse}}{d\,\varepsilon_{axial}} \quad (3)$$

In our numerical calculations, Poisson's ratio is computed using the central finite difference method as [35],

$$v_{xy} = -\frac{\varepsilon_y^{j+1}-\varepsilon_y^{j-1}}{\varepsilon_x^{j+1}-\varepsilon_x^{j-1}}, v_{yx} = -\frac{\varepsilon_x^{j+1}-\varepsilon_x^{j-1}}{\varepsilon_y^{j+1}-\varepsilon_y^{j-1}}, \quad (4)$$

where the integer $j$ represents the strain increment number.

The strain-stress relation is calculated using the method described in the references [37,38], which was designed for three dimensional (3D) material. For a 2D system, the stress calculated from the DFT has to be adjusted since the DFT reported stress is largely underestimated due to averaging force over vacuum space. To avoid this, the stress in this work adopts the force per unit length in the unit of N/m.

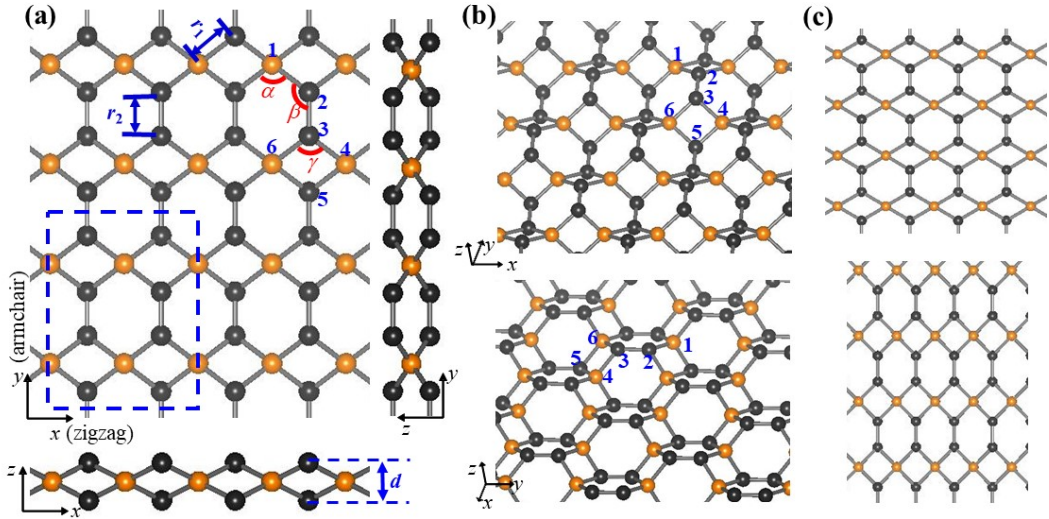

*Figure 1. (a)(b) Snapshots of buckled tetrahex-C. The dashed rectangle represents a unit cell. The $sp^2$ and $sp^3$ hybridized carbon atoms are in black and orange, respectively. (c) Schematics of uniaxial strain in the zigzag (top) and armchair (bottom) directions, respectively.*

## 3. Results and discussion

### A. Structural and mechanical properties

The lattice constants of the relaxed tetrahex-C in our calculations are $a$ = 4.531 Å, $b$ = 6.102 Å, buckling thickness $d$ = 1.163 Å, which are in great agreement with literature [17]. Two distinct bond lengths, three bond angles, and buckling thickness are denoted in Fig. 1(a). Our obtained $r_1$ = 1.534 Å, $r_2$ = 1.338 Å, $\alpha$ = 112.2°, $\beta$ = 123.9°, and $\gamma$ = 95.2°. As shown in Fig.1, two neighbored hexagons in the x-axis are not in a plane, its dihedral angle is denoted as $\phi_{1234}$ determined by the neighboring atoms 1-2-3-4, and our calculated $\phi_{1234}$ = 125.6°. The dihedral angle between two neighbored hexagon and tetragon $\phi_{2345}$ = 137.4°.

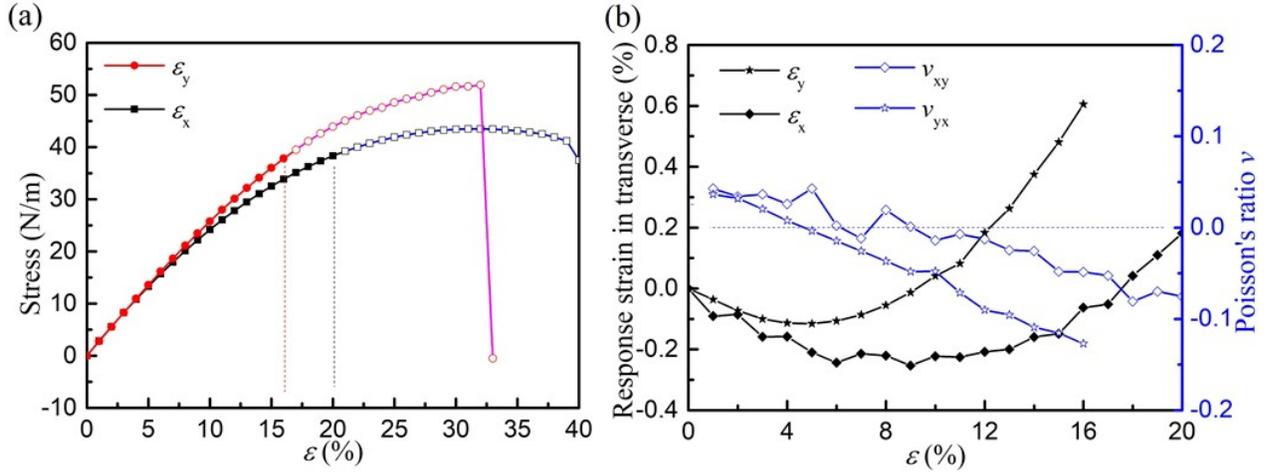

Figure 2. (a) Strain-stress relation in tetrahex-C for uniaxial strain applied in the zigzag (x) and armchair (y) direction, respectively. Phonon instability occurs when strain is beyond 20% (16%) in the x (y) direction with the coressponding strength 38.3 N/m (37.8 N/m). (b) the response strain in the transverse direction and Poisson's ratio. Intrinsice negative Poisson's ratio achieves when axial strain passes a threshold value 7% (5%) in the x (y) direction.

Uniaxial tensile strain is applied to tetrahex-C with an increment of 1% up to 40%. The obtained strain-stress relation is presented in Fig. 2 (a). The material is teared apart with 40% (33%) strain loaded in the x (y) direction. The material is more ductile in the x (zigzag) axis compared to the y (armchair) direction. To check the stability of tetrahex-C under uniaxial tensile strain, the phonon spectrum of the strained structure is calculated. We find that phonon instability occurs near the center of Brillouin zone when strain goes beyond 20% (16%) in the x (y) direction. That leads the ideal strength of tetrahex-C to be 38.3 N/m and 37.8 N/m in the x and y direction, respectively. This ultrahigh strength outperforms penta-graphene which shows 23.5 N/m strength with 18% uniaxial strain in both zigzag and armchair directions [39]. It is comparable to that of graphene, i.e. 36.7 N/m (40.4 N/m) in the zigzag (armchair) direction with phonon instability occurring strain of 19.4% (26.6%) [33]. We re-calculate the strain-stress-relation of graphene using exact same parameters in this work and obtain the ideal strength of 34.4 N/m (38.0 N/m) in the zigzag (armchair) direction.

The response strain in the transverse direction and Poisson's ratio in tetrahex-C are presented in Fig. 2(b). Tetrahex-C shows intrinsic in-plane negative Poisson's ratio when strain goes beyond a threshold value, which is 7% (5%) in the $x$ ($y$) direction. It is found that Poisson's ratio is in the range of -0.081 ~ +0.043 for the uniaxial strain up to 20% in the zigzag direction. Similarly, the value is between -0.127 ~ +0.036 when strain up to 16% is applied in the armchair direction. Poisson's ratio clearly demonstrates anisotropic feature of tetrahex-C.

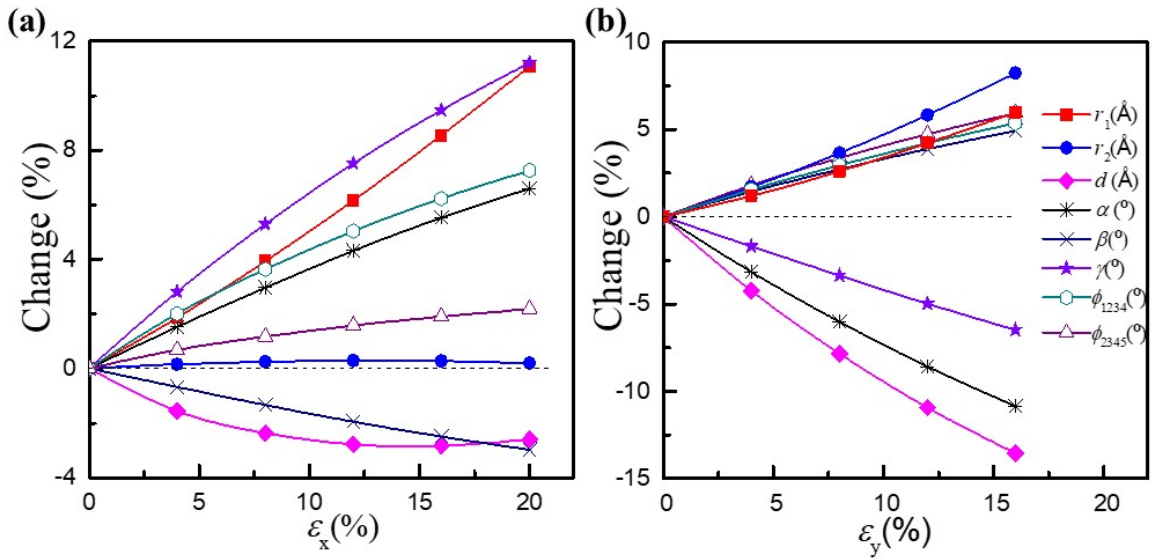

*Figure 3, Structural change of tetrahex-C under uniaxial strain applied in the (a) zigzag and (b) armchair direction. The bond lengths, buckling thickness, bond angles, and dihedral angles are denoted in Figure 1(a). Vertical axis represents the change relative to their original values in the relaxed structure.*

To explore the mechanism of the intrinsic negative Poisson's ratio, the structure geometries of relaxed and strained tetrahex-C are compared and the structural change is presented in Fig. 3. The bond lengths $r_1$, $r_2$, buckling thickness $d$, bond angles $\alpha$, $\beta$ and $\gamma$ are denoted in Fig. 1. $\phi_{1234}$ is the dihedral angle of two neighboring hexagons along the $x$-axis and $\phi_{2345}$ is the dihedral angle of neighboring tetragon and hexgon. The vertical axis in Fig. 3 gives the change of each quantity relative to its original value in the relaxed tetrahex-C. From Fig. 3(a), it is clear that when the structure is under axial strain in the zigzag direction, the leading extension factors are the bond length $r_1$ and bond angle $\gamma$, which both show 11% increase (for the case of $\varepsilon_x$ = 20%) compared to the relaxed structure. Note that the bond length $r_1$ forms the four sides of tetragons and they are all tilted from the x-axis with apparent y-projection. Under the uniaxial strain applied in the x-direction, the expansion in $r_1$ necessarily results in the extension of lattice constant in the y-axis, leading to the negative Poisson's ratio. On the other hand, for the case of strain in the armchair direction in Fig. 3(b), the primary change comes from the lengthening of bond lengths $r_2$ (8.2% increase for the case of $\varepsilon_y$ = 16%) and $r_1$ (6% up), the squeeze of the buckling thickness (13.5% down), and the decrease of bond angle $\alpha$ (10.9% down). The lengthening of $r_1$ and squeezing of buckling thickness $d$ combined together act as the primary factors for the negative Poisson's ratio.

To obtain other mechanical properties, including elastic stiffness constants and moduli, the energy surface of tetrahex-C is scanned in the small strain range -0.6% < $\varepsilon_{xx}$ < +0.6%, -0.6% < $\varepsilon_{yy}$ < +0.6%, and -0.6% < $\varepsilon_{xy}$ < +0.6%. The strain energy is calculated as

$$E_s = E(\varepsilon) - E_0 \tag{5}$$

where $E(\varepsilon)$ and $E_0$ are the total energy of strained and relaxed systems, respectively. The obtained strain energy is then fitted using the following equation,

$$E_s = a_1 \varepsilon_{xx}^2 + a_2 \varepsilon_{yy}^2 + a_3 \varepsilon_{xx} \varepsilon_{yy} + a_4 \varepsilon_{xy}^2 \tag{6}$$

The coefficients $a_i$ in Equation (6) can be determined, and the elastic stiffness constants can be calculated as,

$$C_{ij} = \frac{1}{A_0}\left(\frac{\partial E_s^2}{\partial \varepsilon_i \varepsilon_j}\right) \tag{7}$$

Where $i, j$ = $xx, yy$, or $xy$, $A_0$ is the area of the simulation cell in the $xy$ plane. The Young's and shear moduli for a 2D system can be derived as a function of $a_i$ [22,40],

$$E_x = \frac{4a_1 a_2 - a_3^2}{2a_2 A_0}, E_y = \frac{4a_1 a_2 - a_3^2}{2a_1 A_0}, G_{xy} = \frac{2a_4}{A_0} \tag{8}$$

Our calculated elastic constants in tetrahex-C are $C_{11}$ = 289 N/m, $C_{12}$ = 15 N/m, $C_{22}$ = 282 N/m, $C_{33}$ = 125 N/m, which are in agreement with the literature [17]. The Young's moduli are $E_x$ = 288 N/m and $E_y$ = 281 N/m in the zigzag and armchair directions, respectively. And the calculated shear modulus is $G_{xy}$ = 125 N/m.

Young's and shear moduli along an arbitrary direction can be obtained as following [22],

$$\frac{1}{E_\varphi} = S_{11}\cos^4\varphi + (2S_{12} + S_{66})\cos^2\varphi\sin^2\varphi + S_{22}\sin^4\varphi \tag{9}$$

$$\frac{1}{G_\varphi} = S_{33}(\sin^4\varphi + \cos^4\varphi) + 4\left(S_{11} - 2S_{12} + S_{22} - \frac{1}{2}S_{33}\right)\cos^2\varphi\sin^2\varphi \qquad (10)$$

where $\varphi \in [0, 2\pi]$ is the angle of an arbitrary direction from the +x axis, $E_\varphi$ and $G_\varphi$ are the Young's and shear moduli along that direction, respectively. $S_{ij}$ are elastic compliance constants, which are correlated to elastic stiffness constants,

$$S_{11} = \frac{C_{22}}{C_{11}C_{22} - C_{12}^2}, S_{22} = \frac{C_{11}}{C_{11}C_{22} - C_{12}^2}, S_{12} = -\frac{C_{12}}{C_{11}C_{22} - C_{12}^2}, S_{33} = \frac{1}{C_{33}} \qquad (11)$$

The direction dependence of the Young's and shear moduli is presented in Fig. 4. The maximal Young's modulus is along the x (zigzag) direction with a value of 288 N/m, whereas the minimum 273 N/m is along [11] direction. The average Young's modulus over all directions is 279 N/m. However, it is opposite for the shear modulus. The maximal and minimal shear modulus is along the [11] (135 N/m) and x(y)-direction (125 N/m), respectively.

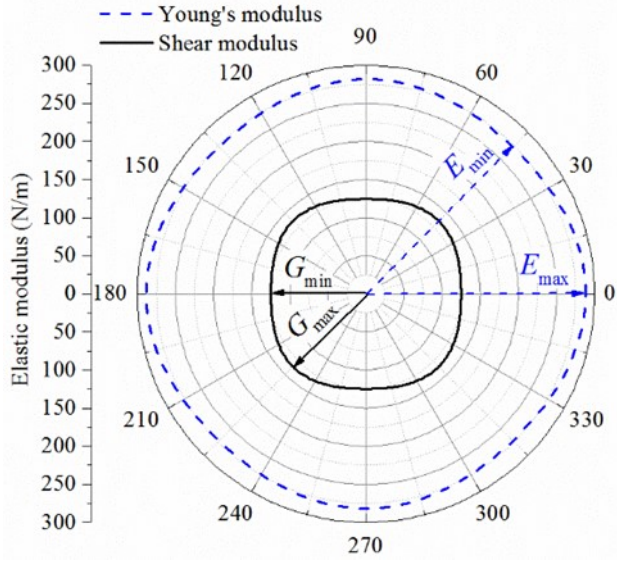

*Figure 4. The directional dependence of Young's (blue dashed line) and shear (black solid line) moduli in tetrahex-C.*

### B. Electronic properties

Tetrahex-C possesses a direct band gap at the center of Brillouin zone [17]. Our calculations confirm this and the band structure is presented in Fig. 5 (a) based on the hybrid HSE functional. Our calculated HSE band gap is 2.64 eV, close to the value of 2.63 eV reported in the reference [17]. Under uniaxial tensile strain in the zigzag direction, the conduction band minimum (CBM, i.e. state C) decreases, while the valence band maximum (VBM, i.e. state B) increases with strain, as shown in Fig. 5(a)-(c), resulting a band gap reduction. The direct-gap feature at Γ remains intact under uniaxial strain in the zigzag direction up to 20% (beyond 20%, phonon spectrum suggests instability of the structure). On the other hand, under axial strain in the armchair direction, the band gap remains direct (or quasi-direct) at Γ till strain up to 16% where phonon instability occurs.

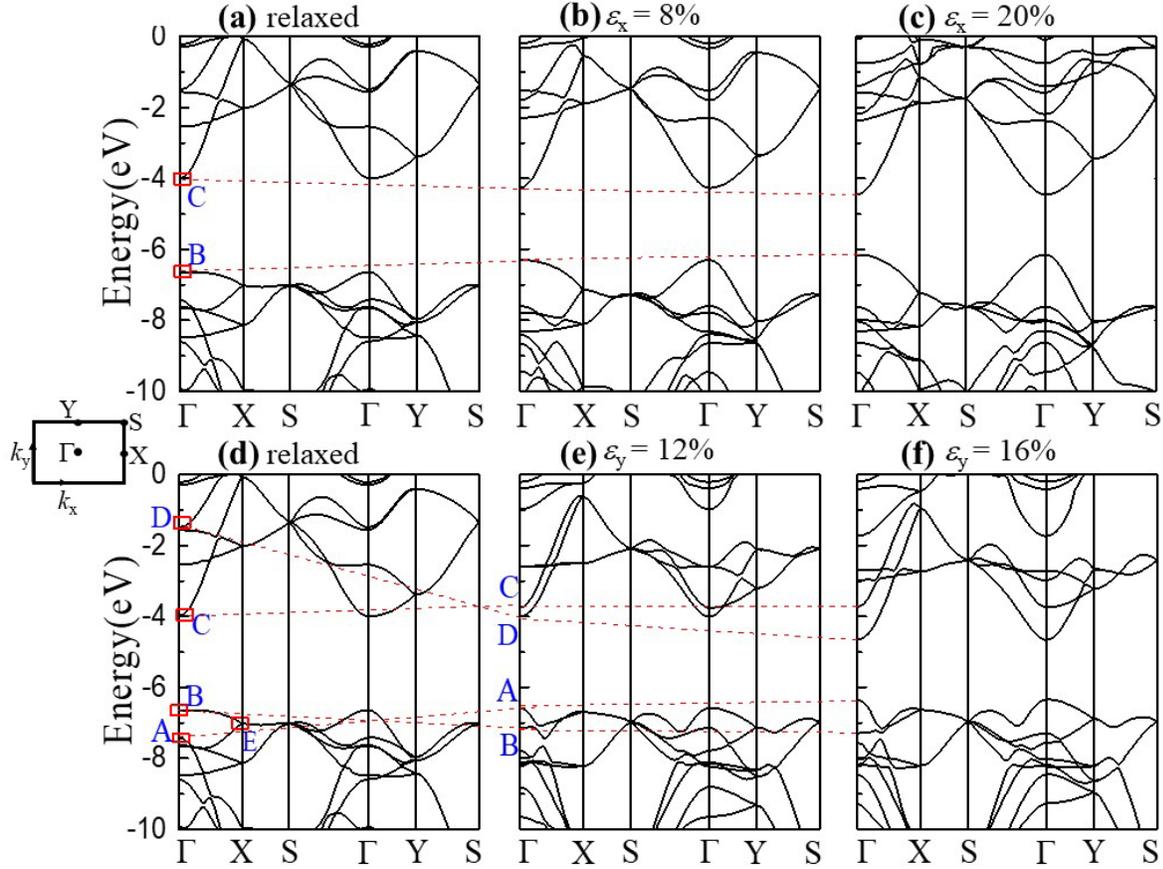

*Figure 5 Variation of electronic band structure with uniaxial strain in tetrahex-C. HSE predicted band structure of (a) relaxed structure, (b) $\varepsilon_x = 8\%$, (c) $\varepsilon_x = 20\%$, (d) relaxed, (e) $\varepsilon_y = 12\%$, and (f) $\varepsilon_y = 16\%$. Energy is referenced to vacuum. The dashed lines are guide for eye for the energy shifts of states A, B, C and D.*

Detailed analysis of the energy variation with strain for the near-edge states (i.e. states A-E labeled in Fig 5) reveals that the CBM remains at state C and VBM stays with state B under strain in the zigzag direction up to 20%. However, energy state crossover takes place for the strain applied in the armchair direction. For example, at $\varepsilon_y = 12\%$ (16%) as shown in Fig. 5(e)-(f), the CBM and VBM are no longer their original states C and B, respectively. Instead, State D (A) becomes the CBM (VBM). The energy of state D decreases rapidly with strain and have a lower energy than state C at $\varepsilon_y = 12\%$ (16%). Similar situation occurs in the valence bands. The energy of State A increases with strain and exceeds that of state B, thus representing the VBM.

We also notice that at $\varepsilon_y = 4\%$ (8%), the VBM slightly shifts away from Γ toward X and the energy difference between the VBM and the Γ point is only 0.03 eV (0.05 eV), leading to tetrahex-C a quasi-direct band gap in the strain range of 4% ~ 9%. The results are consistent with the reference [17] in which the band gap remains direct up to 16.4% of biaxial strain.

The band gap as a function of uniaxial strain in tetrahex-C is presented in Fig. 6(a). Unlike the straight decrease of gap with strain in the $x$ direction, strain in the $y$ direction shows a rich tunability of the gap. To understand this behavior, we further plot the energy variation with strain of the near-edge states A-E in Fig. 6(b)-(d). Fig. 6(b) shows CBM (state C) and VBM (state B) energy with strain $\varepsilon_x$. Both demonstrate relative linear variation which results in a rough liner decrease of band gap with $\varepsilon_x$. However, for strain in the $y$ direction, the VBM experiences an interesting shift from the original state B, to state E, and then to state A at $\varepsilon_y = 9\%$. And CBM moves from state C to state D at $\varepsilon_y = 10.6\%$.

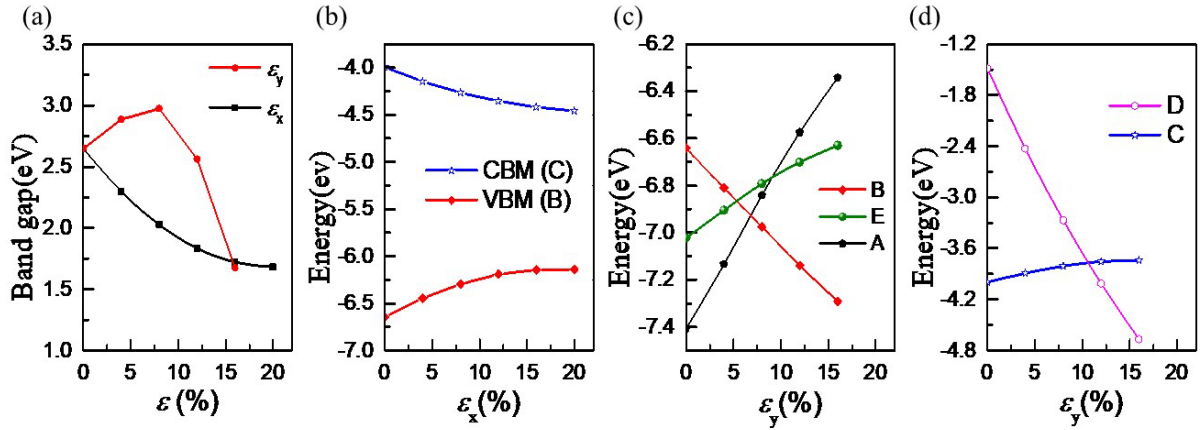

*Figure 6 (a) HSE band gap as a function of uniaxial strain, (b) CBM and VBM energy variation with strain in the zigzag direction, energy variation of (c) valence band states A, B, E, and (d) conduction band states C, D under uniaxial strain in the armchair direction. For valence bands, States B, E energy crossover occurs at 5.4% strain, and states A, E crossover at 9%. For conduction bands, states C, D energy crossover at 10.6% strain.*

The different energy variation patterns of the states in Fig. 6(c)-(d) are related to their specific orbitals and bonding/anti-bonding characteristics. Fig. 7 presents the electron density contour plots of the near-edge states A-E. The wavefunction character of each state is examined by projecting the wavefunction onto $s$-, $p$-, and $d$-orbitals at each ionic site. It is found that state A is contributed by $p_x$-orbitals, while all other states are dominated by $p_z$-orbitals. Examining the dominant *spd*-orbitals along with the phase factors of the wavefunction, one can determine the state's bonding and anti-bonding characteristics along a specific direction. When strain is applied in that specific direction, the energy variation with strain of the state obeys the pattern schematically illustrated in Fig. 7(f) [41,42]. This schematics is derived from the Heitler-London's exchange energy model [43], in which the energies of the bonding and antibonding states are primarily different in terms of the exchange-correlation energy of electrons. The exchange-correlation energy is contributed from either non-classical electron-electron (positive) or electron-ion interaction (negative) with the latter dominant for orbitals with non-localized electron density. Therefore, with applied tensile strain and increase of bond lengths, the energy of bonding states increases while anti-bonding state decreases [41,42].

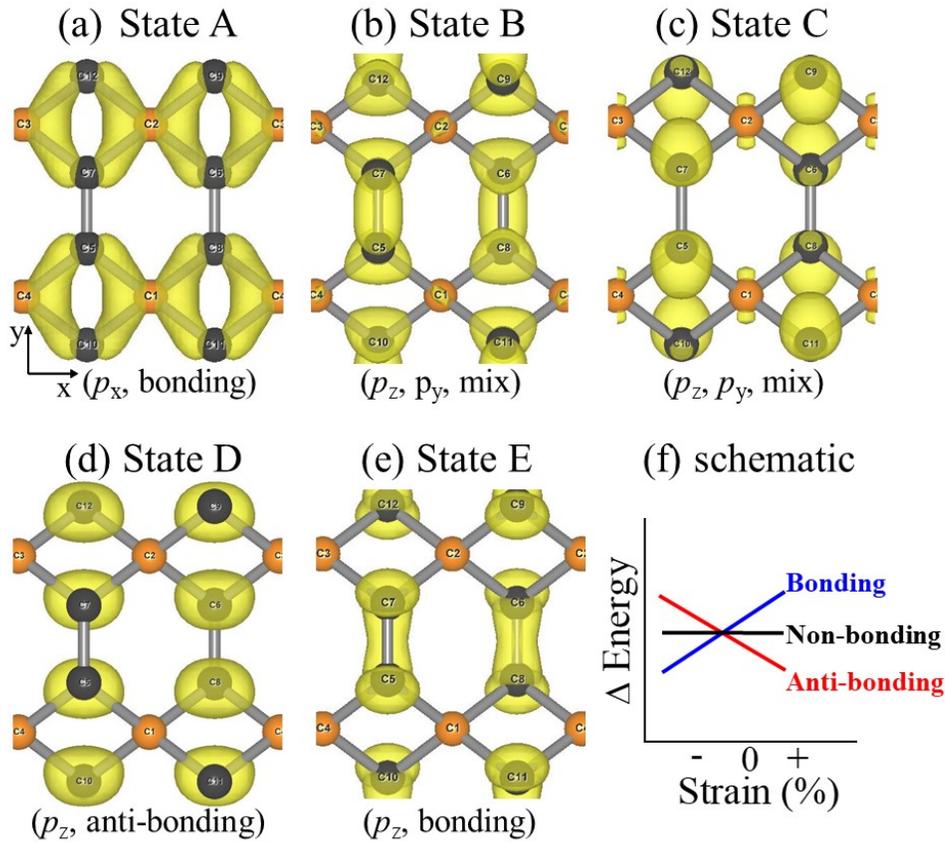

*Figure 7 (a)-(e) the electron density contour plots of the near-band-edge states A - E in tetrahex-C. States A-E are denoted in Figure 5. The isosurface is set to 0.01 e/Å$^3$. Their dominant orbitals and bonding status (examined in the y axis) are listed at the bottom. (f) A schematic of energy response to axial strain for three typical cases of bonding, non-bonding, and anti-bonding.*

From Fig. 7, States A and E possess bonding orbitals in the *y* axis, therefore their energies increase with strain. State D shows anti-bonding nature in the *y*-axis and its energy decreases with strain. State B contributed from 84% $p_z$ and 16% $p_y$-orbitals shows a mix nature with a weak $p_z$ bonding and a strong $p_y$ anti-bonding character along the bond length $r_2$. The overall trend of its energy variation follows the anti-bonding pattern. State C is also a mixture with 76% $p_z$, 19% $p_y$ and 5% $s$-orbitals, in which a weak $p_z$ anti-bonding and a strong $p_y$ bonding characteristics along the bond length $r_2$ result in an overall bonding behavior.

## 4. Summary

We conduct first-principles DFT calculations to investigate the mechanical and electronic properties of tetrahex-C under uniaxial tensile strain along the zigzag and armchair directions, respectively. Tetrahex-C shows ultrahigh strength which outperforms both graphene and penta-graphene. It shows superior ductility and remains phonon stability with uniaxial tensile strain up to 20% (16%) in the zigzag (armchair) direction. This 2D carbon also demonstrates tunable intrinsic negative Poisson's ratio when axial strain is beyond a threshold value of 7% (5%) in the zigzag (armchair) direction. This auxetic property retains in a large strain range of 10% ~ 20% (5% ~ 16%) in the zigzag (armchair) direction. Tetrahex-C has a direct band gap of 2.64 eV (HSE gap) at Γ. This direct-gap feature remains intact upon the entire range of axial strain application up to 20% (16%) in the zigzag (armchair) direction till phonon instability occurs. The band gap is tunable in the range of 1.68 ~ 2.97 eV with uniaxial strain. The ultrahigh strength, negative Poisson's ratio, and direct-band-gap in tetrahex-C imply potential applications in nanomechanics and nanoelectronics.

**Acknowledgement**

This work is financially supported by the Natural Science Foundation of China (Grant No.: 11965005), and the 111 Project (B17035). The authors thank Arizona State University Advanced Computing Center for providing computing resources (Agave Cluster), and the computing facilities at High Performance Computing Center of Xidian University.

* To whom correspondence should be addressed.  E-mail: xihong.peng@asu.edu.